\documentclass[twocolumn,tighten]{aastex62}
\usepackage{amsmath,bm,graphicx,color,capt-of} 

\newcommand{\vv}{\mathbf{v}}
\newcommand{\vB}{\mathbf{B}}
\newcommand{\vJ}{\mathbf{J}}

\newcommand{\bomega}{\boldsymbol{\Omega}}

\newcommand{\grad}{\boldsymbol{\nabla}}
\newcommand{\bcdot}{\boldsymbol{\cdot}}
\newcommand{\dvg}{\boldsymbol{\nabla}\!\bcdot\!}
\newcommand{\curl}{\boldsymbol{\nabla}\!\boldsymbol{\times}\!}
\newcommand{\cnabla}{\!\boldsymbol{\cdot}\!\boldsymbol{\nabla}}
\newcommand{\cross}{\!\!\boldsymbol{\times}\!\!}

\newcommand{\Ro}{\mathrm{Ro}}
\newcommand{\Prt}{\mathrm{Pr}}
\newcommand{\Pm}{\mathrm{Pm}}
\newcommand{\Ren}{\mathrm{Re}}

\newcommand{\rmv}{\mathrm{v}}

\shorttitle{Stellar Dynamos}
\shortauthors{Augustson et al.}

\begin{document}

\title{Rossby and Magnetic Prandtl Number Scaling of Stellar Dynamos}

\correspondingauthor{K.~C. Augustson} 
\email{kyle.augustson@cea.fr}
\author{K.~C. Augustson}
\affil{AIM, CEA, CNRS, Universit\'{e} Paris-Saclay, Universit\'{e} Paris Diderot, Sarbonne Paris Cit\'{e}, F-91191 Gif-sur-Yvette Cedex, France}

\author{A.~S. Brun}
\affil{AIM, CEA, CNRS, Universit\'{e} Paris-Saclay, Universit\'{e} Paris Diderot, Sarbonne Paris Cit\'{e}, F-91191
  Gif-sur-Yvette Cedex, France}

\author{J. Toomre}
\affil{JILA \& Department of Astrophysical and Planetary Sciences, University of Colorado, Boulder, Colorado, USA, 80309}

\begin{abstract}
  Rotational scaling relationships are examined for the degree of equipartition between magnetic and kinetic energies in
  stellar convection zones. These scaling relationships are approached from two paradigms, with first a glance at
  scaling relationship built upon an energy-balance argument and second a look at a force-based scaling. The latter
  implies a transition between a nearly-constant inertial scaling when in the asymptotic limit of minimal diffusion and
  magnetostrophy, whereas the former implies a weaker scaling with convective Rossby number. Both scaling relationships
  are then compared to a suite of 3D convective dynamo simulations with a wide variety of domain geometries,
  stratifications, and range of convective Rossby numbers.
\end{abstract}


\section{Introduction}

Magnetic fields influence both the dynamics and evolution of stars. Detecting such magnetic fields can be difficult
given that the bulk of their energy resides below the surface.  Such magnetic fields are either generated through
convective dynamos, or they are remnant field left over from a star's formation situated in stable radiative zones.  At
some point in their evolution, nearly all stars have convective regions capable of generating magnetic fields.  In
particular, during the pre-main sequence, most stars are fully convective and rapidly rotating, which likely leads to
the development of magnetic fields that are then frozen into the forming radiative regions of these stars.  This implies
that there may be a strong connection between the dynamical magnetic fields in convection zones and the secularly
evolving magnetic fields in radiative regions, unless significant magnetic boundary layers form to prevent an active
codevelopment of the two regions \citep[e.g.,][]{gough98,wood18}.

From the perspective of stellar evolution, the presence of a magnetic field in stable regions can lead to instabilities
and angular momentum transport. Therefore, it is of keen interest to estimate the magnitude of magnetic fields generated
in convective regions that may be radiatively stable in later stages of evolution
\citep{mestel87,spruit02,zahn07,mathis13}.  Doing so is quite challenging due to the inherently nonlinear character of
convective dynamos.  However, such an undertaking has been attempted in \citet{emeriau17}. Alternatively, there exist
several models of convection and its scaling with rotation rate and magnetic field, which can provide an estimate of the
kinetic energy in convective regions. There are also several regimes for which scaling relationships for the ratio of
magnetic energy versus kinetic energy may permit the estimation of the magnitude of the rms magnetic fields, where they
are distinguished by the assumptions about which processes dominate the energetic and force balances.

The effects of astrophysical dynamos can be detected at the surface and in the environment of magnetically-active
objects such as stars \citep[e.g.,][]{christensen09,donati09,donati11,brun15}. One direct way to approximate the
dynamics occurring within such an object is to conduct laboratory experiments with fluids that have some equivalent
global properties, while observing their response to controllable parameters, such as the strength of a thermal forcing
or a rotation rate. In those cases, all the observable scales of the system can be accounted for, from the global or
driving scale to the dissipation scales. In practice, this has proven to be quite difficult when trying to mimic
geophysical or astrophysical dynamos, but they are still fruitful endeavors
\citep[e.g.,][]{gailitis99,laguerre08,spence09}.  However, recent experiments with liquid gallium have shown that
magnetostrophic states, where the Coriolis force balances the Lorentz force, seem to be optimal for heat transport
\citep{king15}, which is interesting given the strong likelihood that many astrophysical dynamos are in such a
state. Another method is to simulate a portion of those experiments, but these numerical simulations are limited in the
scales they can capture: either an attempt is made to resolve a portion of the scales in the inertial range to down to
the physical dissipation scale \citep[e.g.,][]{mininni09a,mininni09b,brandenburg14}, or an attempt is made to
approximate the equations of motion for the global scales while modeling the effects of the unresolved dynamical scales
\citep[e.g.,][]{gilman83,brun04,christensen06,strugarek16,varela17}.

These varying approaches to gathering data about the inner workings of convective dynamos provide a touchstone for
thought experiments. Further, one can attempt to identify a few regimes for which some global-scale aspects of those
dynamos might be estimated with only a knowledge of the basic parameters of the system
\citep[e.g.,][]{christensen10,davidson13,oruba14}. The following questions are examples of such parametric dependencies:
how the magnetic energy contained in the system may change with a modified level of turbulence (or stronger driving), or
how does the ratio of the dissipative length scales impact that energy, or how does rotation influence it?  Establishing
the global-parameter scalings of convective dynamos, particularly with stellar mass and rotation rate, is useful given
that they provide an order of magnitude approximation of the magnetic field strengths generated within the convection
zones of stars as they evolve from the pre-main-sequence to a terminal phase. This in turn permits the placement of
better constraints upon transport processes, such as those for chemical elements and angular momentum, most of which
occur over evolutionary timescales.

\begin{figure}[t]
  \centering
    \includegraphics[width=0.95\linewidth]{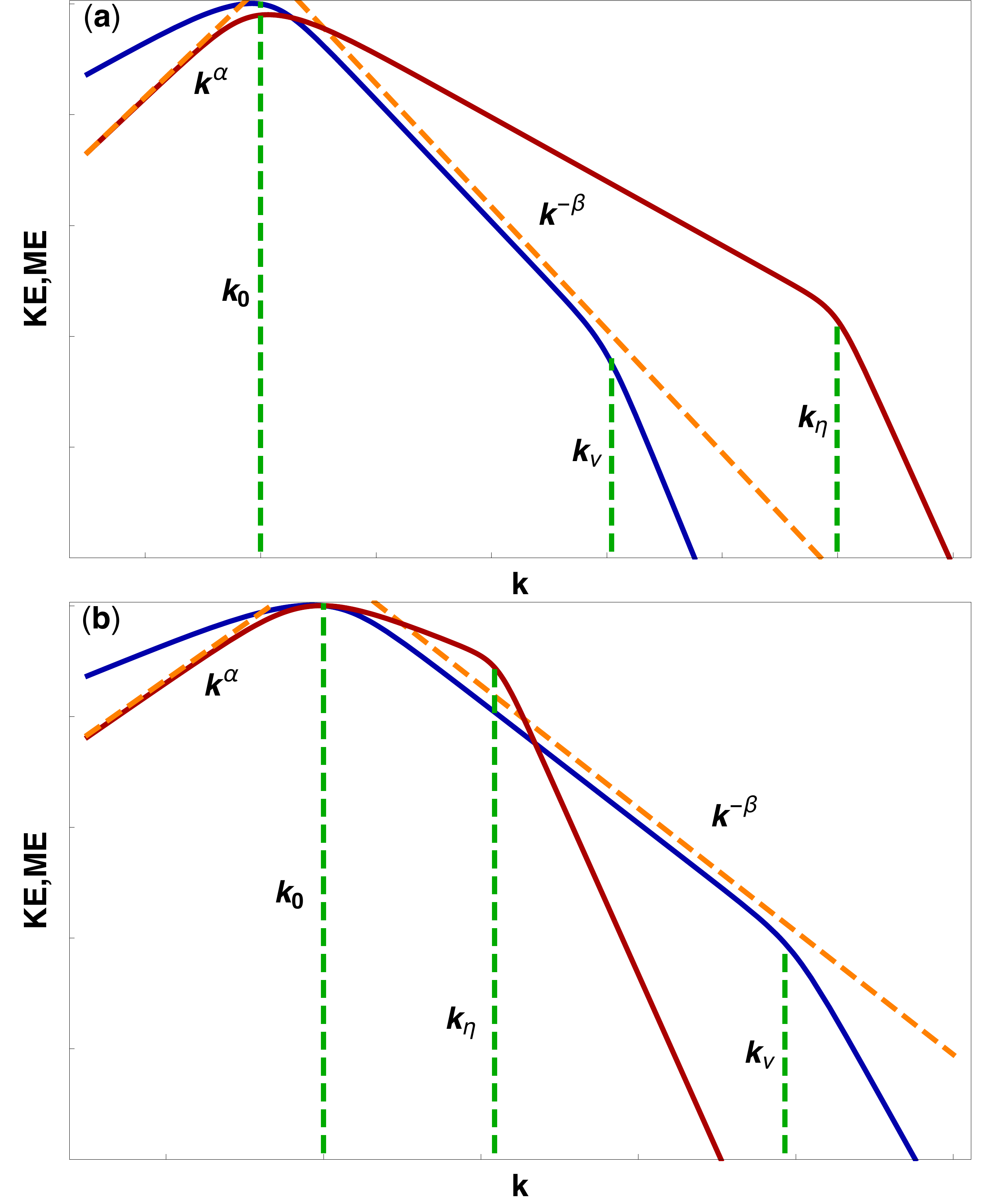}
    \caption{A sketch of energy spectra for low and high-$\Pm$ dynamos with scale wavenumber $k=2\pi/\ell$, with
      magnetic energy in red and kinetic energy in blue showing (a) a saturated high-$\Pm$ dynamo, and (b) a saturated
      low-$\Pm$ dynamo. The dashed vertical green lines characterize distinct wavenumbers: $k_0=2\pi/\ell_0$ is an
      integral scale, $k_{\nu}=2\pi/\ell_\nu$ the viscous dissipation scale, and $k_{\eta}=2\pi/\ell_\eta$ the Joule
      scale. The dashed orange line indicates the small-wavenumber scaling of the magnetic field with
      $\mathrm{ME}\propto k^\alpha$ and the inertial range of the kinetic energy with $\mathrm{KE}\propto
      k^{-\beta}$.} \label{fig:sketch}
\end{figure}

\section{Fundamental Equations}

In the hunt for a simple set of algebraic equations to describe the basic convective and dynamo processes at work in
stellar convection zones, it is useful to consider the following set of MHD equations:

\vspace{-0.25truein}
\begin{center}
  \begin{align}
    &\frac{\partial\rho}{\partial t} = -\dvg{\left(\rho\vv\right)}, \label{eqn:con}\\
    &\rho\frac{\partial\vv}{\partial t} = -\rho\left(\right.\!\!\vv\cnabla{\!\left.\right)\!\vv}
    - 2\rho\bomega\cross\vv - \grad{P} - \rho\grad{\Phi_{\mathrm{eff}}}\nonumber\\
    &+\frac{\vJ\cross\vB}{c} +\grad{\!\bcdot\sigma}, \label{eqn:eom} \\
    &\frac{\partial\vB}{\partial t} = \curl{\left[\vv\cross\vB-\frac{4\pi\eta}{c}\vJ\right]}, \label{eqn:ind}\\
    &\grad{\bcdot\vB}=0, \label{eqn:divB}\\
    &\frac{\partial E}{\partial t} = -\dvg{\left[\left(E + P - \sigma\right)\vv +\mathbf{q}\right]} + H, \label{eqn:eng}
  \end{align}
\end{center}

\noindent where $\vv$ is the velocity, $\vB$ is the magnetic field, $\rho$ is the density, $P$ the pressure, and $s$ is
the entropy per unit mass. Moreover, the following variables are also defined as
$\Phi_{\mathrm{eff}} = \Phi + \lambda^2\Omega^2/2$, $\Phi$ is the gravitational potential, $\Omega$ is the rotation rate
of the frame, $\lambda$ is the distance from the axis of rotation, the current is $\vJ = c\curl{\vB}/4\pi$, with $c$
being the speed of light, $\sigma_{ij}=2\rho\nu\left(\mathbf{e}_{ij}-1/3\dvg{\vv}\delta_{ij}\right)$ is the viscous stress
tensor where $\mathbf{e} = 1/2\left(\grad{\vv}+\grad{\vv}^T\right)$, $\mathbf{q} = -\kappa \nabla T$, $\kappa$ is the
thermal diffusivity and $H$ is an internal heating rate per unit mass that is due to some prescribed exoergic
process (e.g., chemical, nuclear, or otherwise). The total energy is
$E =\rho \rmv^2/2 + B^2/8\pi + \rho\Phi_{\mathrm{eff}} + \rho e$, where $e$ is the internal energy per unit mass. Since
only the most basic scaling behavior of the stellar system is sought, the following assumptions are made: the total
energy of the system is conserved and the system is in a nonlinearly-saturated statistically-stationary equilibrium.

\section{Distinct Magnetic Prandtl Number Regimes}\label{sec:prandtl}

In considering stellar dynamos, the choice of an appropriate magnetic Prandtl number arises: should one consider those
defined by atomic diffusivities or those defined by turbulent diffusion mechanisms. If the former, then the typical
Prandtl number can be quite small. If the latter, then the Prandtl number is typically near unity. Here it will be
considered that the ordering of time-scales is preserved even if the turbulent diffusivities are invoked, therefore the
thermal and magnetic Prandtl numbers will remain fixed at their atomic values thereby preserving the hierarchy of length
and time-scales that they set, which is not without precedence \citep{schekochihin04,brandenburg05}. On the other hand,
with turbulent diffusivities, it would seem that all stellar dynamos would be in the moderate magnetic Prandtl number
regime, where $\Pm\approx 1$.

If the time scale of the flows at a scale $\ell$ is given by $\tau_\ell = \ell/\rmv_\ell$, where $\rmv_\ell$ is the
characteristic velocity at a length scale of $\ell$, and the gradient operators for a magnetic and velocity field of scales
$\ell_B$ and $\ell$ respectively yield $\nabla \vB \approx B_{\ell_B}/\ell_B$ and $\nabla \vv \approx \rmv_\ell/\ell$,
then if the magnitude of each term in Equation \ref{eqn:ind} is considered as

\vspace{-0.25truein}
\begin{center}
  \begin{align}
    \frac{\rmv_\ell B_{\ell_B}}{\ell} &=  2\frac{\rmv_\ell B_{\ell_B}}{\ell} + \frac{\rmv_\ell B_{\ell_B}}{\ell_B} +
                                        \frac{\eta B_{\ell_B}}{\ell_B^2}.
  \end{align}
\end{center}

\noindent Therefore, assuming that the hierarchy of scales is preserved along the turbulent inertial range and defining the scale
Reynolds number as $\Ren_\ell = \rmv_\ell \ell/\nu$, a reformulation of the above equation yields

\vspace{-0.25truein}
\begin{center}
  \begin{align}
    &\frac{\ell_B^2}{\ell^2} \Pm \Ren_\ell\left( 1+\frac{\ell}{\ell_B}\right) \approx 1,
  \end{align}
\end{center}

\noindent where $\Pm=\nu/\eta$ is the magnetic Prandtl number, which implies that $\ell_B/\ell \approx \Pm^{-1/2}$ as
seen in both the kinematic and nonlinear regimes of time-dependent ABC-flow dynamos \citep{arnold83,brummell01}.
Therefore, at large magnetic Prandtl number, the length scales of the magnetic field are smaller than that of the
velocity field. At low $\Pm$, the expectation should be the opposite that the generated magnetic fields roughly tend to
be of larger scale than the velocity field. This does not preclude smaller scale intermittent fields for the low-$\Pm$
regime or vice versa, rather it is more a statement about which scales of the magnetic field contain the most energy:
the large-scales for low-$\Pm$ dynamos and the small-scales for high-$\Pm$ dynamos as is sketched in Figure
\ref{fig:sketch} and seen in nonlinear 3D dynamo simulations \citep{brummell01,schekochihin04,brandenburg05}.

Using the Braginskii definitions of the viscous and magnetic diffusivities (see Appendix \ref{app2}) and turning to
stellar models, one can determine the ranges of magnetic Prandtl numbers within a star. For dynamos in stars with
convective envelopes, the low-$\Pm$ regime is of greatest interest. Using the Sun as an example, the atomic $\Pm$
ranges between about $10^{-1}$ in the core to roughly $10^{-6}$ at the photosphere, decreasing exponentially toward
it. Other low mass stars are roughly similar, with a low $\Pm$ throughout the bulk of their interiors. High mass stars,
in contrast, have relatively large atomic values of $\Pm$ ranging from about $10$ throughout the convective core and
remaining nearly unity until the near-surface region is reached at which point it drops to about $10^{-2}$.

\begin{figure}[t]
  \centering
  \includegraphics[width=0.95\linewidth]{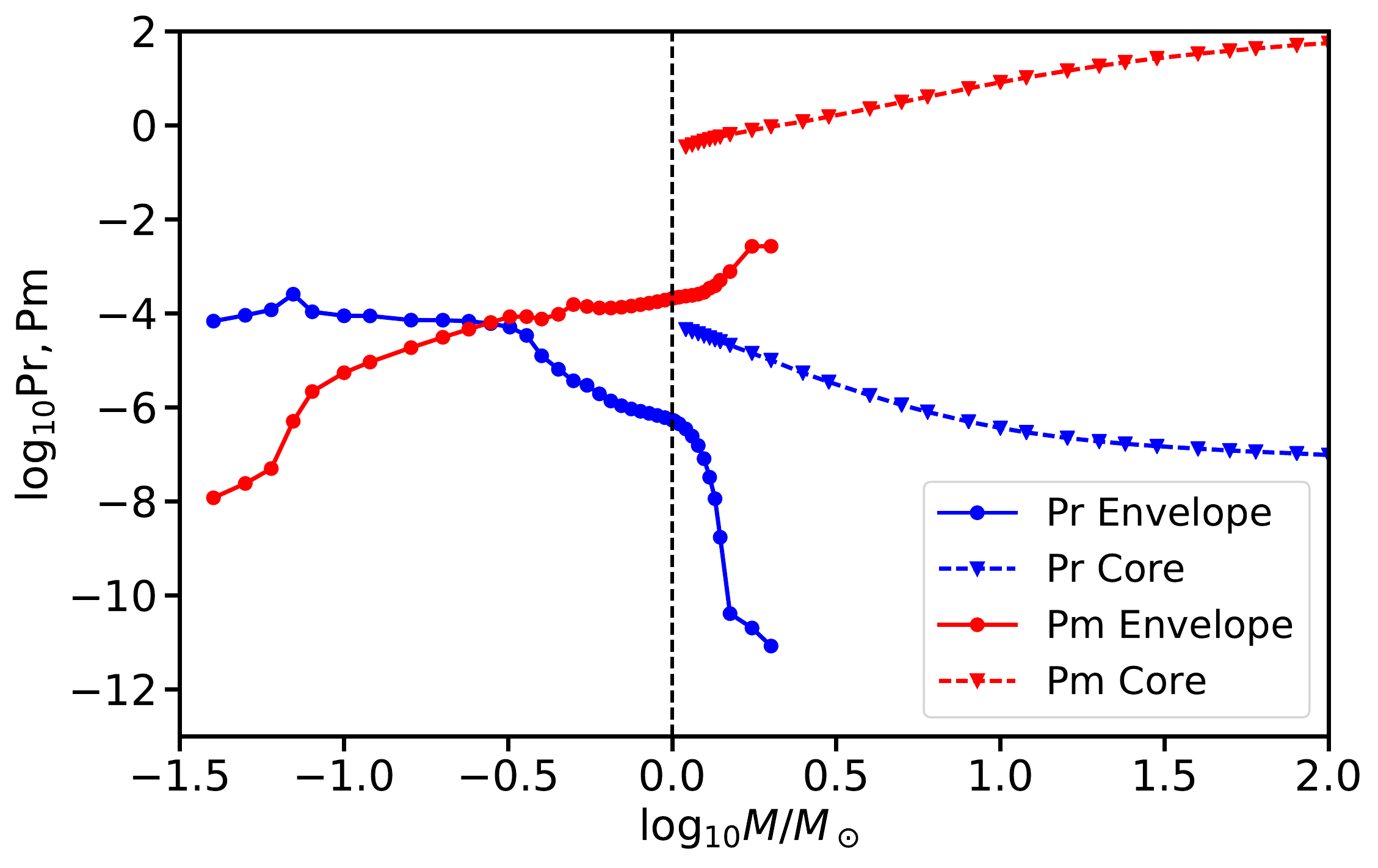}
  \caption{The average thermal Prandtl number $\Prt$ and average magnetic Prandtl number $\Pm$ for ZAMS stars with
    masses between 0.03 and 100~$M_{\odot}$, with the average being taken over substantial convection zones.  The zone
    being averaged is indicated by triangles for a convective core and circles for a convective envelope. }
  \label{fig:prandtl}
\end{figure}

The prescription for the atomic values of the Prandtl numbers given in Appendix \ref{app1} has been applied to models of
stars near the zero-age main-sequence with the solar metallicity in the mass range between 0.03 and 100 $M_{\odot}$
computed with MESA \citep{paxton11,paxton13,paxton15,paxton18}.  The resulting convective-zone-averaged Prandtl numbers
are shown in Figure \ref{fig:prandtl}, where it is clear that all stars possess convective regions with a low thermal
Prandtl number ($\Prt=\nu/\kappa$, as seen in Figure \ref{fig:prandtl}a), whereas one can find two regimes of magnetic
Prandtl number (Figure \ref{fig:prandtl}b). The existence of these two regimes is directly related to where the
convective region is located. For massive stars with convective cores, the temperature and density averaged over the
convective volume are quite high when compared to lower-mass stars with a convective envelope. Such a high temperature
leads to a large magnetic Prandtl number.  The dichotomy in magnetic Prandtl number implies that there may be two
fundamentally different kinds of convective dynamo action in low-mass versus high-mass stars.

\section{The Energetic Balance}\label{sec:energybalance}

Of the convective dynamo topics, it is the geodynamo that has received the most focused theoretical attention
\citep[e.g.,][]{buffett00,roberts00,buffett09}, but it also has a unique niche of its own not shared by most stellar
dynamos.  The Rossby number of mantle flows is estimated to be approximately $10^{-6}$ \citep{finlay11,aurnou15}, which
implies that the convective overturning times are of the order of several thousand years since the Earth rotates in one
day.  Therefore, the estimated convective velocities are of order $1 \mathrm{km\, yr^{-1}}$ in the mantle.  Thus,
Coriolis forces take center stage, and one may be justified in ignoring the nonlinear advection of vorticity though not
of heat or composition.  Primordial magnetism is ruled out by ohmic losses that could dissipate core magnetism on time
scales of order $10^4 \mathrm{yr}$, and magnetic reversals in the recent past of about $10^5 \mathrm{yr}$.

The recent spherical shell simulations of the geodynamo have focused on trying to minimize the Ekman number
$\mathrm{Ek}$ (the ratio of viscous to Coriolis forces) to as small values as computationally feasible, and to examine
the resulting force balances achieved and the scaling relationships that may result.  These are discussed in overview for two of
the groups of modeling efforts \citep[e.g.,][]{christensen06,king13,oruba14,schrinner14,yadav16}. The guiding notion is
that convective motions in the real geodynamo should be strongly influenced by the resulting magnetic fields, within a
so-called MAC balance involving magnetic (M, or Lorentz), buoyancy (A, for Archimedean), and Coriolis forces, with
inertial forces and viscous stresses having a negligible role.  However, the current hero simulations attaining
$\mathrm{Ek}$ of order $10^{-7}$ still do not attain a clear MAC balance, as discussed by \citet{hughes16} and
\citet{cattaneo17} in offering other limiting equations that may be used to explore weak and strong magnetic field
states \citep[e.g.,][]{dormy16}.

In the spirit of the MAC balance, one approach to building a scaling relationship for the ratio of magnetic to kinetic
energy considers the balances established in generating entropy, kinetic energy, and magnetic energy as well as the
force balance. To begin, note that the evolution of the magnetic energy is

\vspace{-0.25truein}
\begin{center}
  \begin{align}
    &\frac{\partial}{\partial t}\left(\frac{\vB^2}{8\pi}\right) =
    \frac{\vB}{4\pi}\!\bcdot\!\curl{\left[\vv\cross\vB-\frac{4\pi\eta}{c}\vJ\right]}, \nonumber \\
    &=-\frac{1}{c}\left[\dvg{\left(\eta\vJ\cross\vB\right)}+\frac{4\pi\eta}{c}\vJ^2+\vv\!\bcdot\!\left(\vJ\cross\vB\right)\right].
  \end{align}
\end{center}

\noindent If this equation is averaged over many dynamical times $\tau$, when it is in a quasi-steady state, and if it
is integrated over the volume of the convective region, it yields

\vspace{-0.25truein}
\begin{center}
  \begin{align}
    &\!\!\int\!\!\frac{dt}{\tau} dV\left[\frac{4\pi\eta}{c}\vJ^2+\vv\!\bcdot\!\left(\vJ\cross\vB\right)\right] = -\!\!\int\!\!\frac{dt}{\tau} d\mathbf{S}\!\bcdot\!\left(\eta\vJ\cross\vB\right). \label{eqn:intme}
  \end{align}
\end{center}

\noindent The Poynting flux of the right-hand-side of Equation \ref{eqn:intme} can vanish at the boundaries of the
convective region for an appropriate choice of boundary conditions.  As an example, if the magnetic field satisfies a
potential field boundary condition, then it is zero. Or if the field is force-free (e.g., $\vJ\propto\vB$), then it is
also zero.  Supposing that this is the case, then one has that the average Lorentz work
($\int\!\!dt/\tau dV\vv\!\bcdot\!\left(\vJ\cross\vB\right)$) is equal to the average Joule heating
$H_{\eta}=4\pi\!\!\int\!\! dt/\tau dV\eta\vJ^2/c$.  This is an important point for it shows that the nature of the
convection and magnetic field structures are directly impacted by the form of the resistive dissipation.  Hence, the use
of numerical dissipation schemes could yield unexpected results and the boundary conditions may leak magnetic energy
through a Poynting flux unless they are chosen carefully.

In a fashion similar to that used for the magnetic energy evolution above, one can find that the kinetic energy evolves
as

\vspace{-0.25truein}
\begin{center}
  \begin{align}
    \frac{1}{2}\frac{\partial\rho\vv^2}{\partial t} &=
    -\dvg{\left[\left(\frac{1}{2}\rho\vv^2+P-\sigma\right)\vv\right]} + P\dvg{\vv} - \sigma\!:\!\grad\vv \nonumber \\
    &-\rho\vv\cnabla\Phi_{\mathrm{eff}} + \frac{\vv}{c}\!\bcdot\!\left(\vJ\cross\vB\right).
  \end{align}
\end{center}

\noindent To eliminate the compressibility term, the total internal energy must also be added to the system as

\vspace{-0.25truein}
\begin{center}
  \begin{align}
    &\frac{\partial}{\partial t}\left[\frac{1}{2}\rho\vv^2 + \frac{\vB^2}{8\pi} + \rho e\right] = \nonumber \\
    &-\dvg{\left[\left(\frac{1}{2}\rho\vv^2 + \rho e + P - \sigma\right)\vv - \frac{\eta}{c}\vJ\cross\vB\right]} \nonumber \\
    &-\rho\vv\cnabla{\Phi_{\mathrm{eff}}} -\frac{4\pi\eta}{c^2}\vJ^2 -\sigma\!:\!\grad\vv + H-\dvg{\mathbf{q}}.
  \end{align}
\end{center}

One can also consider the time-averaged and volume-integrated evolution equation for this
energy equation, which yields

\vspace{-0.25truein}
\begin{center}
  \begin{align}
    &\!\!\int\!\! \frac{dt}{\tau} dV\left[H-\dvg{\mathbf{q}}
      -\rho\vv\cnabla\Phi_{\mathrm{eff}}-\frac{4\pi\eta}{c^2}\vJ^2-\sigma:\grad\vv\right]=\nonumber \\
    &\!\!\int\!\! \frac{dt}{\tau}
    d\mathbf{S}\!\bcdot\!\left[\left(\frac{1}{2}\rho\vv^2 + \rho e + P - \sigma\right)\vv - \frac{\eta}{c}\vJ\cross\vB\right].
  \end{align}
\end{center}

\noindent The surface integral is zero if there are no outflows or net torque from the Lorentz force at the domain
boundaries, implying the following:

\vspace{-0.25truein}
\begin{center}
  \begin{align}
    &\!\!\int\!\! \frac{dt}{\tau} dV\left[H-\dvg{\mathbf{q}}
      -\rho\vv\cnabla\Phi_{\mathrm{eff}}-\frac{4\pi\eta}{c^2}\vJ^2-\sigma:\grad\vv\right]=0. \label{eqn:nrg}
  \end{align}
\end{center}

\noindent Note that $\int\! dV H = L(r)$, where $L$ is the total luminosity of the star at a given radius $r$
for a spherically symmetric heating.  Likewise, the radiative luminosity of the star is given by
$\int\! dV \grad\!\!\bcdot\!{\mathbf{q}} = L_r(r) = -4\pi r^2\kappa\partial T/\partial r$ for the spherically-symmetric
component of the temperature field, which should be dominant. For the case of stars that are on the main-sequence, there
are three configurations of their primary regions of convection: either a convective core for high mass stars, a
convective envelope for lower mass stars, or both for F-type stars (see Figure \ref{fig:prandtl}).  In these cases,
one can assume that the region of integration is over the entire convection zone and so the volume-integrated luminosity
will be the nuclear luminosity, or the current total luminosity $L_{*}=L(R_{*})$.  Furthermore, the radiative luminosity
will be nearly, but not exactly, equal in magnitude to $L_{*}$.  The reason for this is that the thermal evolution of
the system is a largely passive response to the changes in the nuclear burning rates.  Because the nuclear luminosity is
slowly increasing along the main-sequence, $L_R=L_r(R_{*})$ will always lag behind $L_{*}$ due to the time required for
thermal diffusion to modify the thermal gradient. From Equation (\ref{eqn:nrg}), one can see that

\vspace{-0.25truein}
\begin{center}
  \begin{align}
    L_{*}\!-\! L_R \!+\! H_\nu \!+\! H_\eta &\!=\! -\!\!\!\int \!\!\frac{dt}{\tau} dV \rho\vv\cnabla\Phi_{\mathrm{eff}} \!=\! W_B, \label{eqn:fullnrg}
  \end{align}
\end{center}

\noindent where $H_\nu$ and $H_\eta$ are the positive-definite, time-averaged, volume-integrated dissipation rates due
to viscosity and resistivity, respectively. Thus, the rate of buoyancy work $W_B$ is directly proportional to the
mismatch of the two luminosities and the rates of viscous and resistive dissipation, implying that the latter result
from the former. 

Returning to the time-averaged curl of the momentum equation, though neglecting the viscous and inertial terms under the
assumption that the Rossby number is sufficiently small and that the viscous terms play no role, one can find

\vspace{-0.25truein}
\begin{center}
  \begin{align}
    \curl{\left[2\rho\vv\cross\bomega+\frac{1}{c}\vJ\cross\vB\right]}+\grad\rho\cross\mathbf{g}_{\mathrm{eff}}=0, \label{eqn:curlmom}
  \end{align}
\end{center}

\noindent where $\mathbf{g}_{\mathrm{eff}}=-\grad\Phi_{\mathrm{eff}}$. This provides the basis of finding the length
scales of the flows and the magnetic fields. Neglecting the inertial terms does mean that the domain of applicable
Rossby numbers is restricted to being below unity, whereas neglecting the viscous terms is consistent with assumption of
a low magnetic Prandtl number as discussed in \citet{davidson13}. Specifically, the Rossby number is assumed to be small
enough so that the flow becomes roughly columnar and moderately aligned with the rotation axis. In such a case, there
are two integral length scales of the flow: one parallel to the rotation axis $\ell_{\|}$ and another perpendicular to
it $\ell_{\perp}$, with $\ell_{\perp}<\ell_{\|}$.  These can be defined as
$\ell_{\|} = \left(\langle \rmv^2\rangle/\langle\omega_z^2\rangle\right)^{1/2}$ and
$\ell_{\perp} = \left(\langle \rmv^2\rangle/\langle\omega^2-\omega_z^2\rangle\right)^{1/2}$, where
$\boldsymbol{\omega} = \curl{\vv}$, $\omega_z = \mathbf{\hat{z}}\cnabla\mathbf{\times}\vv$, and $\langle\rangle$ denotes
a volume average. Here, unlike \citet{davidson13}, the density stratification is retained. Therefore, for comparison
purposes, note that the integral scale of the flow is then
$\ell_0 = \left(\langle\rmv^2\rangle/\langle\omega^2\rangle\right)^{1/2} =
\left[\ell_\perp^2\ell_{\|}^2/\left(\ell_{\perp}^2+ \ell_{\|}^2\right)\right]^{1/2}$. Moreover, the density
perturbations rather than temperature perturbations are contained in the buoyancy work integral and the force balance
below, which permits the treatment of the baroclinic flows present in the convection zones of most stars. The full
density can be retained in the integral and in the scaling given that the gradient of the mean density is parallel to
$\mathbf{g}_{\mathrm{eff}}$. So, their cross product vanishes, leaving only the product of the velocity and gradients of
the density perturbations. Assuming further that the magnetic energy density per unit mass scales only with $\ell_{\|}$,
since $\ell_{\perp}<\ell_{\|}$, and $w_B$, unit consistency requires that
$B^2/(4\pi\rho) \propto F\left(\ell_{\|}, w_B\right) \approx \ell_{\|}^{2/3} w_B^{2/3}$, where $w_B$ is the rate of
buoyancy work per unit mass $W_B/M_{\mathrm{CZ}}$ and $M_{\mathrm{CZ}}$ is the mass contained in the convection
zone. Therefore, given Equation (\ref{eqn:curlmom}) and assuming that each term is of the same order of magnitude, the
basic proportionality is

\vspace{-0.25truein}
\begin{center}
  \begin{align}
    &\rho\bomega\cnabla\vv\approx\grad\rho\cross\mathbf{g}_{\mathrm{eff}}\approx\frac{1}{c}\curl{\left(\vJ\cross\vB\right)}\nonumber \\
    \implies &\frac{\Omega \rmv_{\mathrm{rms}}}{\ell_{\|}} \approx \frac{g}{\ell_{\perp}} \approx \frac{B^2}{4\pi\rho\ell_{\perp}^2}, \label{eqn:forcebal}
  \end{align}
\end{center}

\noindent where $\bomega\cnabla$ selects the parallel length scale $\ell_{\|}$, whereas
$\nabla \propto \ell_{\perp}^{-1}$ since $\ell_\perp < \ell_{\|}$. Thus, comparing the curl of the Lorentz force to the
curl of the Coriolis force, one has

\vspace{-0.25truein}
\begin{center}
  \begin{align}
    \frac{\Omega \rmv_{\mathrm{rms}}}{\ell_{\|}} \approx \frac{B^2}{4\pi\rho\ell_{\perp}^2} \approx \frac{\ell_{\|}^{\frac{2}{3}}w_B^{\frac{2}{3}}}{\ell_{\perp}^2},
  \end{align}
\end{center}

\noindent and moreover it can be shown that an estimate of the rms velocity is

\vspace{-0.25truein}
\begin{center}
  \begin{align}
    \rmv_{\mathrm{rms}}\approx \frac{\ell_{\|}^{\frac{5}{3}}w_B^{\frac{2}{3}}}{\ell_{\perp}^2\Omega}.\label{eqn:vrmswb}
  \end{align}
\end{center}

\noindent Within the context of such estimates, the buoyancy term can be written as

\vspace{-0.25truein}
\begin{center}
  \begin{align}
    w_B = \frac{\int \frac{dt}{\tau} dV \rho\vv\bcdot\mathbf{g}_{\mathrm{eff}}}{\int \frac{dt}{\tau}dV\rho} \approx g\rmv_{\mathrm{rms}},
  \end{align}
\end{center}

\noindent which implies that the estimated magnitude for the curl of the buoyancy force in Equation
(\ref{eqn:forcebal}) is $g/\ell_{\perp} \approx w_B/(\rmv_{\mathrm{rms}}\ell_{\perp})$. Then, it is easily
seen that

\vspace{-0.25truein}
\begin{center}
  \begin{align}
    \frac{\ell_{\perp}}{\ell_{\|}} \approx \frac{w_B}{\Omega\rmv_{\mathrm{rms}}^2}.
  \end{align}
\end{center}

\noindent So, the ratio of integral length scales should vary as

\vspace{-0.25truein}
\begin{center}
  \begin{align}
    \frac{\ell_{\perp}}{\ell_{\|}} \approx \frac{w_B^{\frac{1}{9}}}{\Omega^{\frac{1}{3}} \ell_{\|}^{\frac{2}{9}}} = \left(\frac{w_B}{\Omega^3\ell_{\|}^2}\right)^{\frac{1}{9}}.
  \end{align}
\end{center}

\noindent Likewise the ratio of magnetic to kinetic energy then scales as

\vspace{-0.25truein}
\begin{center}
  \begin{align}
    \mathrm{\frac{ME}{KE}} \approx \frac{B^2}{8\pi\left(\frac{1}{2}\rho v_{\mathrm{rms}}^2\right)} \approx \left(\frac{w_B}{\Omega^3\ell_{\|}^2}\right)^{-\frac{2}{9}}.
  \end{align}
\end{center}

\noindent The Rossby number may then be defined as

\vspace{-0.25truein}
\begin{center}
  \begin{align}
    \mathrm{Ro} = \frac{\rmv_{\mathrm{rms}}}{\Omega\ell_{\|}}\approx\left(\frac{w_B}{\Omega^3\ell_{\|}^2}\right)^{\frac{4}{9}}, \label{eqn:rossby}
  \end{align}
\end{center}

\noindent which implies that

\vspace{-0.25truein}
\begin{center}
  \begin{align}
    \mathrm{\frac{ME}{KE}} \propto \mathrm{Ro}^{-\frac{1}{2}}. \label{eqn:buoyancyscaling}
  \end{align}
\end{center}

\noindent To make contact with stellar parameters, the buoyancy work may be estimated noting that
$W_B \approx G M_{*} M_{\mathrm{CZ}} \rmv_{\mathrm{rms}} /R_{\mathrm{CZ}}^2$, where $R_{\mathrm{CZ}}$ is the radius at
which $\rho(R_{\mathrm{CZ}})=\rho_{\mathrm{CZ}}$ and $\rho_{\mathrm{CZ}}$ is the average density of the convection zone.
However, if one takes into account the scaling of the buoyancy work with Rossby number and magnetic Prandtl number,
Equation (\ref{eqn:rossby}) becomes an implicit relationship for the Rossby number that may be indeterminate for large
magnetic Prandtl number (see \S\ref{sec:dis} and \citet{davidson13}).  This can be partially seen through Equation
\ref{eqn:vrmswb} may be recast to solve for $w_B$ as

\vspace{-0.25truein}
\begin{center}
  \begin{align}
    w_B \approx \left(\frac{\ell_{\perp}^2\Omega \rmv_{\mathrm{rms}}}{\ell_{\|}^{\frac{5}{3}}}\right)^{\frac{3}{2}}
    = \rmv_{\mathrm{rms}}^{2}\Omega^{\frac{3}{2}}\frac{\langle\omega_z^2\rangle^{\frac{5}{4}}}{\langle\omega^2-\omega_z^2\rangle^{\frac{3}{2}}}.
  \end{align}
\end{center}

One simple method to estimate these quantities is to suppose that the energy containing flows have roughly the same
length scale as the depth of the convection zone and that the speed of the flows is directly related to the rate of
energy injection (given here by the stellar luminosity) and inversely proportional to the density of the medium into
which that energy is being injected \citep{augustson12,brun17}. This is encapsulated as
$\rmv_{\mathrm{rms}} \propto \left[L_*/\left(2\pi \rho_{\mathrm{CZ}} R_{\mathrm{CZ}}^2\right)\right]^{1/3}$, where
$\rho_{\mathrm{CZ}}$ is the average density in the convection zone and $R_{\mathrm{CZ}}$ is the radius corresponding to
this average density. Therefore, the buoyancy work scales as

\vspace{-0.25truein}
\begin{center}
  \begin{align}
    w_B \approx f \left(\frac{L_*\Omega^{\frac{9}{2}}}{2\pi\rho_{\mathrm{CZ}} R_{\mathrm{CZ}}^2}\right)^{\frac{1}{3}},\label{eqn:wbest}
  \end{align}
\end{center}

\noindent where $f=\langle\omega_z^2\rangle^{\frac{5}{4}}/\langle\omega^2-\omega_z^2\rangle^{\frac{3}{2}}$ is a flow
asymmetry parameter. However, such a mixing-length velocity prescription only provides an order of magnitude estimate as
it too will depend upon the level of turbulence and more fundamentally on the rotation rate
\citep[e.g.,][]{brun17,featherstone16,augustson18a,augustson18b}.

\section{A Force-Balance Scaling} \label{sec:forcescaling}

In addition to assessing the energetic balances of a dynamo, another way to assess the scaling behavior of the magnetic
and kinetic energies in a dynamo is to find the balance of forces acting in the system when in a statistically-steady,
but nonlinear regime.  Before proceeding into the more delicate multi-term balances, and as a means of estimating the
relative order of magnitude of each term that may contribute to the amplitude of the magnetic field, consider first the
four pairwise Lorentz force balances in the nonlinearly-saturated statistically-stationary momentum equation:

\vspace{-0.25truein}
\begin{center}
  \begin{align}
    \frac{\vJ\cross\vB}{c} &\approx \rho \vv\cnabla{\!\vv} \implies \frac{B_{\ell_B}^2}{4\pi\ell_B} \approx \frac{\rho v_\ell^2}{\ell},\label{eqn:si}\\
    \frac{\vJ\cross\vB}{c} &\approx 2\rho\bomega\cross\vv \implies \frac{B_{\ell_B}^2}{4\pi\ell_B} \approx \rho\Omega v_\ell,\label{eqn:sc}\\ 
    \frac{\vJ\cross\vB}{c} &\approx \grad{P'} + \rho' \grad{\Phi_{\mathrm{eff}}} \implies \frac{B_{\ell_B}^2}{4\pi\ell_B} \approx \epsilon \left(\frac{\overline{P}}{\ell} + \overline{\rho} g_{\mathrm{eff}}\right),\label{eqn:sp}\\
    \frac{\vJ\cross\vB}{c} &\approx \grad\bcdot\sigma \implies \frac{B_{\ell_B}^2}{4\pi\ell_B} \approx \frac{\nu \rmv_\ell}{\ell^2},\label{eqn:sv}
  \end{align}
\end{center}

\noindent where the hydrostatic balance has been removed, and for a flow at the scale $\ell$, for which there is an
equivalent magnetic scale $\ell_B$ as discussed in \S \ref{sec:prandtl}.

Convective flows often possess distributions of length scales and speeds that are peaked near a single characteristic
value. Thus, assuming a single characteristic scale for the magnetic and velocity fields, where $\ell_B = \ell=\ell_0$
for which $\rmv_\ell=\rmv_{\mathrm{rms}}$ as in the previous section, for Equations \ref{eqn:si}-\ref{eqn:sv} under the
anelastic approximation this respectively leads to

\vspace{-0.25truein}
\begin{center}
  \begin{align}
    \left.\mathrm{\frac{ME}{KE}}\right|_{I,\mathrm{rms}} &\propto 1,\\
    \left.\mathrm{\frac{ME}{KE}}\right|_{C, \mathrm{rms}} &\propto \Ro_{\mathrm{rms}}^{-1}\\ 
    \left.\mathrm{\frac{ME}{KE}}\right|_{P, \mathrm{rms}} &\propto \frac{\epsilon}{\mathrm{KE}_{\mathrm{rms}}} \left(\overline{P} + \overline{\rho} g_{\mathrm{eff}}\ell\right),\\
    \left.\mathrm{\frac{ME}{KE}}\right|_{V, \mathrm{rms}} &\propto \Ren_{\mathrm{rms}}^{-1},
  \end{align}
\end{center}

\noindent where the subscript $I$ represents the inertial balance, $C$ the magnetostrophic balance, $P$ the buoyancy and
pressure work balance, and $V$ the viscous balance. In anelastic systems $\epsilon \approx \mathrm{Ma}^2$, where
$\mathrm{Ma}$ is the Mach number, implying that the potential energy represented by
$\overline{P} + \overline{\rho}g_{\mathrm{eff}}\ell$ is much smaller than the other possible balances for the bulk of
most stellar convection zones since $\mathrm{Ma}\ll1$ for most stars except near their photosphere.  Since stars are
often rotating fairly rapidly, taking for instance young low-mass stars and most intermediate and high-mass stars, their
dynamos may reach a quasi-magnetostrophic state wherein the Coriolis acceleration also plays a significant part in
balancing the Lorentz force.  Such a balance has been addressed and discussed at length in \citet{christensen10} and
\citet{brun15} for example.

To better characterize the force balance, consider Equation (\ref{eqn:eom}) while taking its curl, wherein one can see
that

\vspace{-0.25truein}
\begin{center}
  \begin{align}
    \frac{\partial\boldsymbol{\omega}}{\partial t} &= \curl{\left[\vv\cross\boldsymbol{\omega_P}+\frac{1}{\rho}\left(\frac{\vJ\cross\vB}{c}+\dvg{\sigma}-\grad P\right)\right]},
  \end{align}
\end{center}

\noindent where $\boldsymbol{\omega} = \curl{\vv}$ and $\boldsymbol{\omega_P}=2\bomega+\boldsymbol{\omega}$.  

Taking the dot product of this equation with $\boldsymbol{\omega}$ gives rise to the equation for
the evolution of the enstrophy.  Integrating that equation over the volume of the convective domain
and over a reasonable number of dynamical times such that the system is statistically steady yields

\vspace{-0.25truein} 
\begin{center}
  \begin{align}
    &\!\!\int\!\! d\mathbf{S}\!\bcdot\!\left[\vv\cross\boldsymbol{\omega_P}+\frac{1}{\rho}\left(\frac{\vJ\cross\vB}{c}+\dvg{\sigma}-\grad P\right)\right]\cross\boldsymbol{\omega} \nonumber \\
    &\!\!+\!\!\int\!\! dV \left(\curl{\boldsymbol{\omega}}\right)\!\bcdot\!\left[\vv\cross\boldsymbol{\omega_P}+\frac{1}{\rho}\left(\frac{\vJ\cross\vB}{c}+\dvg{\sigma}-\grad P\right)\right]\!=\!0.
  \end{align}
\end{center}

\noindent If no enstrophy is lost through the boundaries of the convective domains, then the surface
integral vanishes, leaving

\vspace{-0.25truein} 
\begin{center}
  \begin{align}
    \!\!\int\!\! dV \left(\curl{\boldsymbol{\omega}}\right)\!\bcdot\!\left[\vv\cross\boldsymbol{\omega_P}+\frac{1}{\rho}\left(\frac{\vJ\cross\vB}{c}+\dvg{\sigma}-\grad P\right)\right]=0.
  \end{align}
\end{center}

\noindent This assumption effectively means that magnetic stellar winds will not be part of this scaling analysis.  For
time scales consistent with the dynamical time scales of the dynamos considered here, this is a reasonably valid
assumption. Then, since $\boldsymbol{\nabla\!\times\!\omega}$ is not everywhere zero, the terms in square brackets must
be zero, which implies that

\vspace{-0.25truein} 
\begin{center}
  \begin{align}
    \rho\vv\cross\boldsymbol{\omega_P}+\frac{\vJ\cross\vB}{c}+\dvg{\sigma}-\grad P=0.
  \end{align}
\end{center}

\noindent Taking the curl of this equation eliminates the pressure contribution and gives

\vspace{-0.25truein}
\begin{center}
  \begin{align}
    \curl{\left[\rho\vv\cross\boldsymbol{\omega}+2\rho\vv\cross\bomega+\frac{\vJ\cross\vB}{c}+\dvg{\sigma}\right]}=0.\label{eqn:curlforce}
  \end{align}
\end{center}

\noindent This is the primary force balance, being between inertial, Coriolis, Lorentz, and viscous forces.  Here the
buoyancy force vanishes through the first curl, and the pressure force through the second one.  However, another choice
could be made where the density dependence is retained, leaving a baroclinic term that does not appear in earlier dynamo
considerations where the flows are typically barotropic. Taking fiducial values for the parameters, and scaling the
derivatives as the inverse of a characteristic velocity length scale $\ell$, the scaling relationship for the above
equation yields

\vspace{-0.25truein} 
\begin{center}
  \begin{align}
    \frac{\rho \mathrm{v_{rms}^2}}{\ell^2} + \frac{2\rho \mathrm{v_{rms}}\Omega}{\ell} + \frac{B^2}{4\pi\ell^2} + \frac{\rho\nu \mathrm{v_{rms}}}{\ell^3} \approx 0.
  \end{align}
\end{center}

\noindent If it is assumed that the flows are self-similar at each scale, as in the Kolmogorov turbulence model
\citep{kolmogorov41}, then ratio of the velocity length scale to the magnetic length scale follows from the argument in
\S\ref{sec:prandtl} as $\ell/\ell_B = \Pm^{1/2}$.  In this case, when divided through by
$\rho\mathrm{v_{rms}^2}/\ell^2$, the previous equation simplifies to

\vspace{-0.25truein} 
\begin{center}
  \begin{align}
    \mathrm{\frac{ME}{KE\,\Pm}} \propto 1 + \mathrm{Re}^{-1} + \Ro^{-1}.
  \end{align}
\end{center}

\noindent The Reynolds number used in the above equation is taken to be $\mathrm{Re} = \mathrm{v_{rms}}\ell/\nu$.  Since
the curl is taken, the approximation for the pressure gradient and buoyancy terms employed in \citet{augustson16} is
eliminated from the force balance.  However, the leading term of this scaling relationship is found to be less than
unity, at least when assessed through simulations.  Hence, it should be replaced with a parameter to account for dynamos
that are subequipartition, leaving the following

\vspace{-0.25truein} 
\begin{center}
  \begin{align}
    \mathrm{\frac{ME}{KE\, \Pm}} \propto a + b\mathrm{Ro}^{-1}, \label{eqn:forcescaling}
  \end{align}
\end{center}

\noindent where $a$ and $b$ are unknown a priori as they depend upon the intrinsic ability of the rotating system to
generate magnetic fields, which in turn depends upon the specific details of the system such as the boundary conditions
and geometry of the convection zone. Instead, they must be fit for currently as is discussed below. As demonstrated in
\citet{augustson16}, Equation \ref{eqn:forcescaling} may hold for a subset of convective dynamos, wherein the ratio of
the total magnetic energy (ME) to the kinetic energy (KE) depends on the inverse Rossby number and a constant offset.
The constant is sensitive to details of the dynamics and, in some circumstances, it may also be influenced by the Rossby
number. In any case, convective dynamos are sensitive to the degree of the rotational constraint on the convection, as
it has a direct impact on the intrinsic ability of the convection to generate a sustained dynamo.  Yet, even in the
absence of rotation, there appears to be dynamo action that gives rise to a minimum magnetic energy state in the case of
sufficient levels of turbulence.  Hence, there is a bridge between two dynamo regimes: the equipartition slowly rotating
dynamos and the rapidly rotating magnetostrophic regime, where $\mathrm{ME/KE}\propto \Ro^{-1}$.  For low Rossby
numbers, or large rotation rates, it is even possible that the dynamo can reach superequipartition states where
$\mathrm{ME/KE}>1$. Indeed, it may be much greater than unity, as is expected for the Earth's dynamo (see Figure 6 of
\citet{roberts13}).

\begin{figure}[!t]
  \centering
  \includegraphics[width=0.95\linewidth]{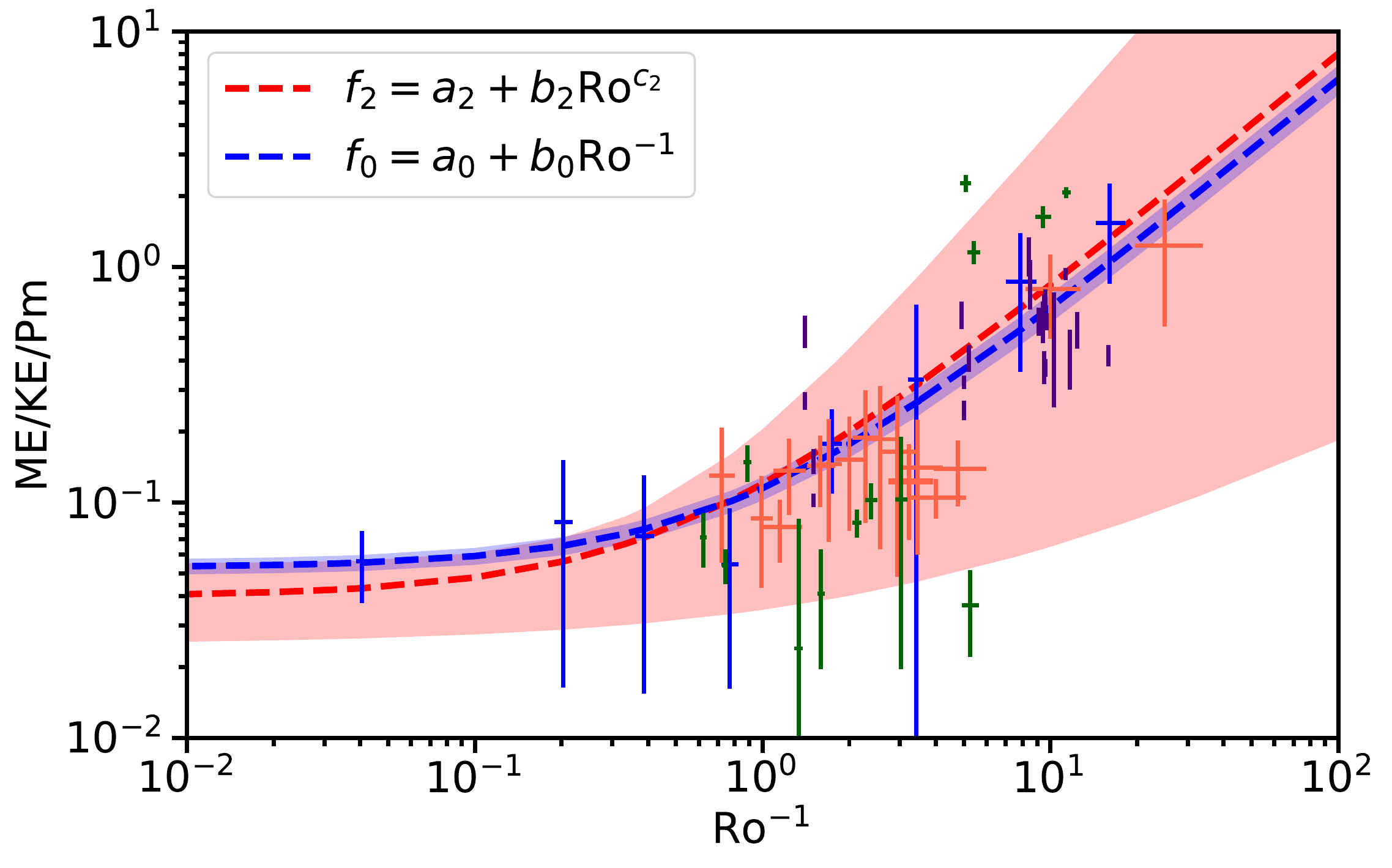}
  \caption{The scaling of the ratio of magnetic to kinetic energy with inverse Rossby number ($\mathrm{Ro}^{-1}$) for
    the restricted data set. The dashed red curve indicates the scaling $f_2$ defined in Equation (\ref{eqn:genRofit})
    with the mean parameters $a_2=0.04\pm0.02$, $b_2=0.08\pm0.04$, and $c_2=-1.0\pm0.4$.  The blue dashed line is for
    the force balance scaling of Equation (\ref{eqn:forcefit}), with the mean parameters $a_0=0.053\pm0.004$ and
    $b_0=0.062\pm0.009$. The shaded regions correspond to the uncertainty bounds of those coefficients. The uncertainty
    of the measured Rossby number and energy ratio that arises from temporal variations are indicated by the size of the
    cross for each data point, with data in blue from \citet{augustson16}, in green from \citet{varela16}, in indigo
    from \citet{kapyla17}, and in orange from \citet{strugarek18}.}
  \label{fig:scaling}
\end{figure}

\begin{figure}[!t]
  \centering
  \includegraphics[width=0.95\linewidth]{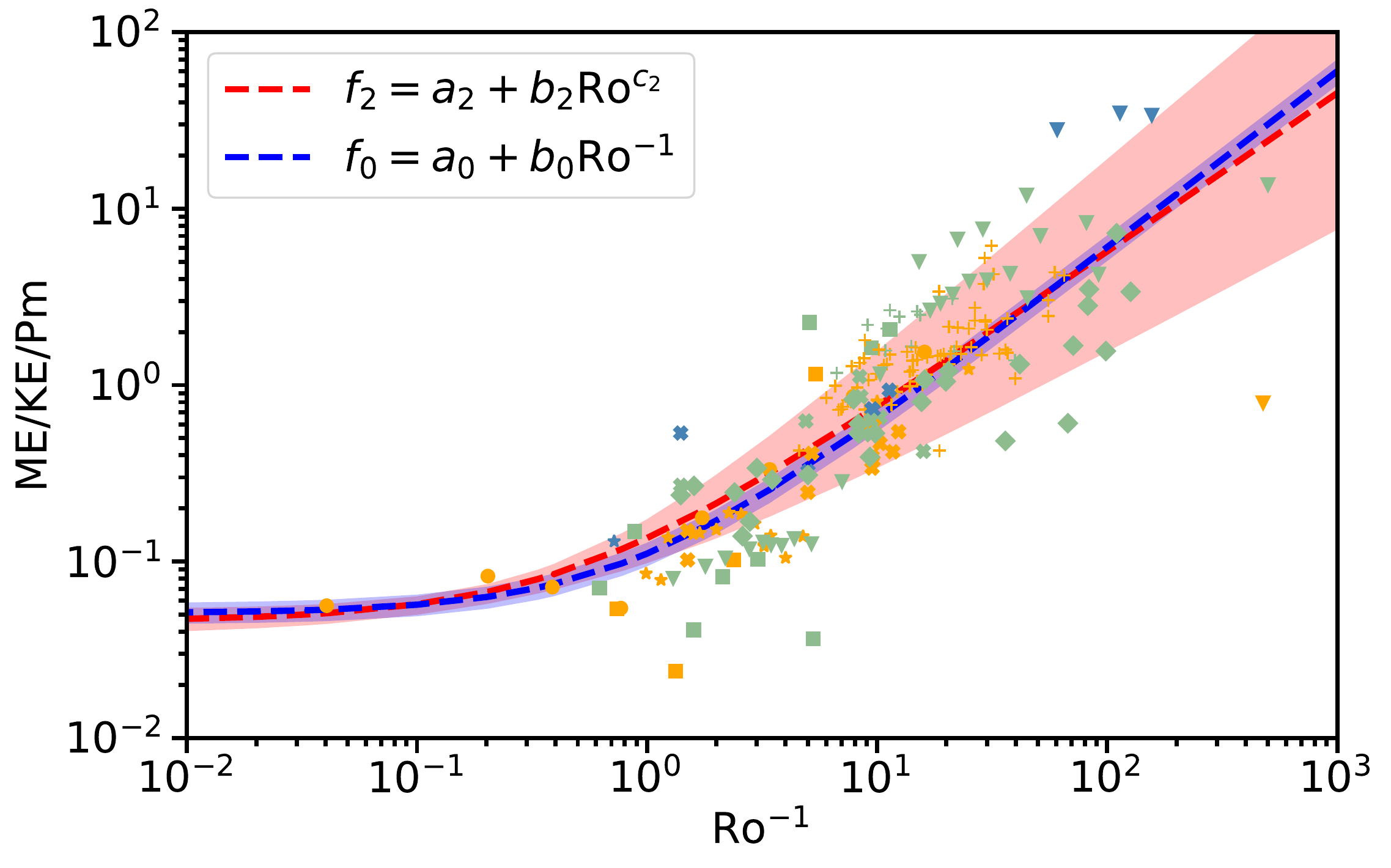}
  \caption{The scaling of the ratio of magnetic to kinetic energy with inverse Rossby number ($\mathrm{Ro}^{-1}$) for
    the full data set. The color of the data point indicates whether its magnetic Prandtl number ($\mathrm{Pm}$) is less
    than unity (light blue), unity (light green), or greater than unity (gold).  The symbols indicate the data source
    with plus from \citet{schrinner14}, circles from \citet{augustson16}, squares from \citet{varela16}, triangles from
    \citet{yadav16}, crosses from \citet{kapyla17}, stars from \citet{strugarek18}, and diamonds from
    \citet{viviani18}. The dashed red curve indicates the scaling $f_2$ defined in Equation (\ref{eqn:genRofit}) with
    the mean parameters $a_2=0.046\pm0.008$, $b_2=0.09\pm0.03$, and $c_2=-0.9\pm0.2$.  The blue dashed line is for the
    force balance scaling of Equation (\ref{eqn:forcefit}), with the mean parameters $a_0=0.053\pm0.007$ and
    $b_0=0.062\pm0.01$. The shaded regions correspond to the uncertainty bounds in those coefficients, which are given
    in Table \ref{tab1}.}
  \label{fig:scalingfull}
\end{figure}

\section{Discussion}\label{sec:dis}

To assess which models provide an accurate description of the physics underlying convective dynamo action, first
consider four increasingly general fits to the data with

\vspace{-0.25truein} 
\begin{center}
  \begin{align}
    f_0 &= a_0 + b_0 \Ro^{-1},\label{eqn:forcefit}\\
    f_1 &= a_1 \Ro^{b_1} \Pm^{c_1},\\
    f_2 &= a_2 + b_2 \Ro^{c_2},\label{eqn:genRofit}\\
    f_3 &= a_3 + b_3 \Pm^{c_3} + d_3 \Ro^{e_3} + f_3 \Ro^{g_3} \Pm^{h_3},
  \end{align}
\end{center}

\noindent where the latter function is effectively the first order multivariate expansion of
$\mathrm{ME/\left(KE Pm\right)}$.

Following \citet{stelzer13}, the fitting procedure to determine the coefficients and their variance is the bounded
nonlinear least-squares technique in log space, which provides one measure of the goodness of an individual fit as

\vspace{-0.25truein} 
\begin{center}
  \begin{align}
    \chi_j^2 = \frac{\ln{10}}{n}\sum_{i=1}^n \left(\frac{\log_{10}\left(y_i\right)-\log_{10}\left(f_{j}\left(\Ro_i,\Pm_i\right)\right)}{\sigma_i/y_i}\right)^2,
  \end{align}
\end{center}

\noindent where $y_i$ are the data $\mathrm{ME/\left(KE Pm\right)}$ for a given $\Ro_i$ and $\Pm_i$, and where $\sigma_i$ are the
standard deviations of the temporal variance of the data, which can be determined for four sets of data from
\citet{augustson16,varela16,kapyla17}, and \citet{strugarek18}.  In order to ascertain the goodness of fit globally, the
fitting procedure is applied to $n$ subsets of the data, where each subset is constructed by removing the $i$th data
point, following the cross-validation method. The cross-validation procedure yields a measure of the goodness of fit
as 

\vspace{-0.25truein} 
\begin{center}
  \begin{align}
    \chi_{\mathrm{CV},j}^2 = \frac{\ln{10}}{n}\sum_{i=1}^n \left(\frac{\log_{10}\left(y_i\right)-\log_{10}\left(f_{j}^*\left(\Ro_i,\Pm_i\right)\right)}{\sigma_i/y_i}\right)^2,
  \end{align}
\end{center}

\noindent where the model function $f_j^*$ is constructed using the subset of data excluding the $i$th point. Finally,
an additional measure of goodness of fit is the relative misfit defined as

\vspace{-0.25truein} 
\begin{center}
  \begin{align}
    \chi_{\mathrm{rel},j} = \sqrt{\frac{1}{n}\sum_{i=1}^n \left(\frac{y_i-f_{j}\left(\Ro_i,\Pm_i\right)}{y_i}\right)^2},
  \end{align}
\end{center}

\noindent which again may be applied to each entire data set and within the cross-validation scheme as

\vspace{-0.25truein} 
\begin{center}
  \begin{align}
    \chi_{\mathrm{rel},CV,j} = \sqrt{\frac{1}{n}\sum_{i=1}^n \left(\frac{y_i-f_{j}^*\left(\Ro_i,\Pm_i\right)}{y_i}\right)^2}.
  \end{align}
\end{center}

\noindent Note further that, since each data subset for the cross-validation yields a different set of fitting
parameters, one may construct a probability distribution for each of those parameters.

\begin{figure*}[!t]
  \centering
  \includegraphics[width=0.95\linewidth]{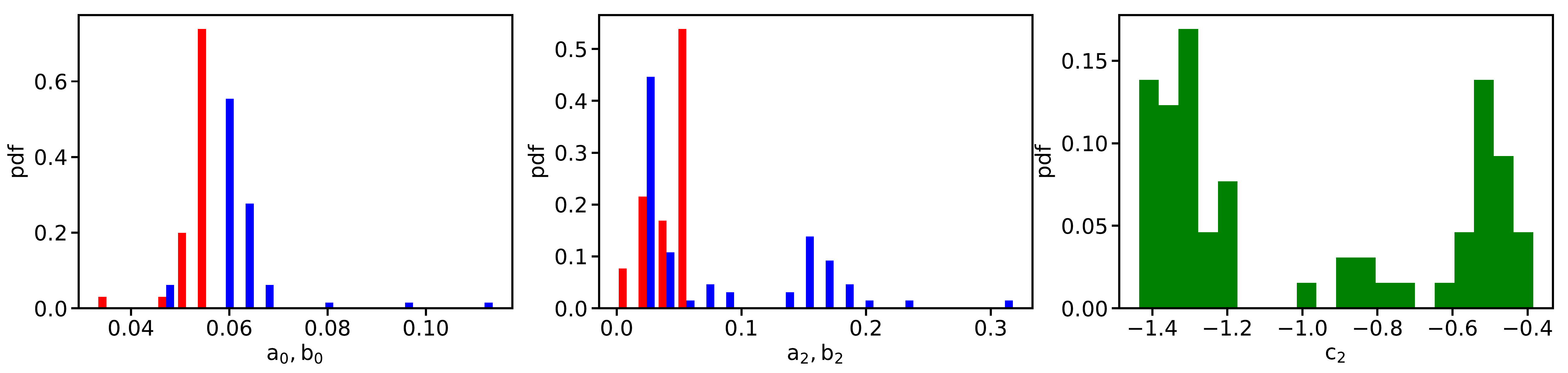}
  \caption{A histogram of the distribution of fitting parameters from the cross-validation analysis for the data in Figure
    \ref{fig:scaling} for the fits of $f_0$ ($a_0$ (red) and $b_0$ (blue) in the leftmost panel) and $f_2$ ($a_2$ (red) and $b_2$
    (blue) in the middle panel, and $c_2$ in the right panel).}
  \label{fig:fit2}
\end{figure*}

The fits resulting from the cross-validation method applied to data set with known variance are shown alongside the data
and its variances in Figure \ref{fig:scaling}, with the caveat that the Rossby number variance for the data is not given
in \citet{kapyla17}. For the full data set, which includes data from
\citet{schrinner14,augustson16,kapyla17,varela17,strugarek18}, and \citet{viviani18}, corresponding to the data shown in
Figure \ref{fig:scalingfull}. The fit parameters and their uncertainties for both data sets are detailed in Table
\ref{tab1}.  Since the variance of the Rossby number and magnetic to kinetic energy ratio was not available for the full
data set, it is set to be a constant equal to the average variance of the cases that do possess this measurement. Only
the statistically significant fits are shown in Figures \ref{fig:scaling} and \ref{fig:scalingfull}, with the
determination of significance made from the goodness of fit parameters given in Table \ref{tab2}.

To consider the applicability of the force-balance and energy-based scaling regimes discussed in the previous sections,
the magnetic Prandtl number and Rossby number dependence of the ratio of the magnetic and kinetic energy has been
assessed in seven sets of simulation data. Five sets of this data are from 3D spherical MHD stellar simulations given in
Tables 1 and 2 respectively in \citet{augustson16} for core dynamo simulations of massive stars, \citet{varela16} for
low-mass stars with a stable region, \citet{strugarek18} for Sun-like stars with an outer convective layer, and
\citet{kapyla17} as well as \citet{viviani18} also for Sun-like stars. Although one may also qualitatively consider the
simulations presented in \citet{mabuchi15}, \citet{guerrero16}, and \citet{warnecke18} for differential rotation and
dynamo action in Sun-like stars that cover a range of rotation rates and magnetic Prandtl numbers, the data provided in
those papers are not sufficient to compare them quantitatively to the other simulations. Specifically, the trends of the
magnetic to kinetic energy ratio seen in those papers are consistent with those determined here. Unpublished data that
was used to construct the data shown in those papers was reanalyzed to determine the temporal variability of the data
from \citet{augustson16} and \citet{varela16}.  To extend this analysis into the domain of geodynamo simulations, data
from a suite of 3D spherical MHD geodynamo simulations is also included from \citet{schrinner14} Table C.1 and the
online data provided in \citet{yadav16}.  These simulations range from being Boussinesq to highly-stratified with both
convectively stable and unstable regions.  The bulk of these simulations have magnetic Prandtl numbers that are close to
unity, with $\Pm$ being typically between 0.25 and 10.

\begin{figure*}[!t]
  \centering
  \includegraphics[width=0.95\linewidth]{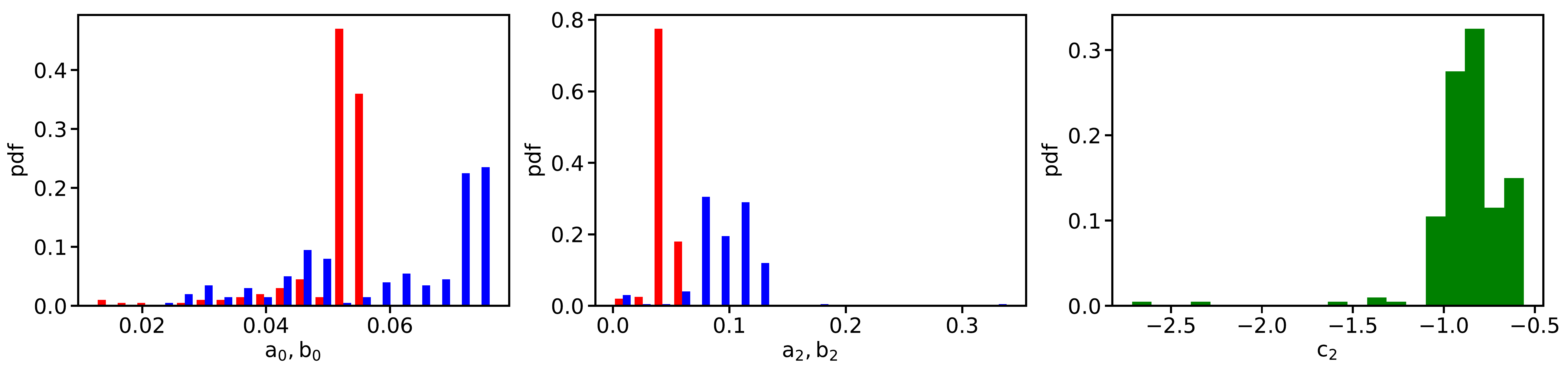}
  \caption{A histogram of the distribution of fitting parameters from the cross-validation analysis for the data in Figure
    \ref{fig:scalingfull} for the fits of $f_0$ ($a_0$ (red) and $b_0$ (blue) in the leftmost panel) and $f_2$ ($a_2$ (red) and $b_2$
    (blue) in the middle panel, and $c_2$ in the right panel).}
  \label{fig:fit4}
\end{figure*}

With this data in hand, one may compute the best-fit to the data for each of the fitting functions described above, with
the statistically significant fits for Equations \ref{eqn:forcefit} and \ref{eqn:genRofit} shown in Figures
\ref{fig:scaling} and \ref{fig:scalingfull}. Histograms of the distributions of these parameters are shown for the
statistically relevant fits in Figures \ref{fig:fit2} and \ref{fig:fit4}.  In Figure \ref{fig:fit2} most parameters have
monomodal distributions.  However, it would seem that there are two populations for the fitting parameter $c_2$, yet
this is most likely an artifact given that the distribution for $c_2$ in Figure \ref{fig:fit4} is no longer strongly
bimodal. The fitting functions that include a magnetic Prandtl number dependence, with the coefficients given in Tables
\ref{tab1} and \ref{tab2}, are not well constrained given the scatter in the data with respect to both magnetic Reynolds
number and magnetic Prandtl number. Hence, the best that can be said about the scaling of the ratio of the magnetic
energy to kinetic energy is with respect to the general trend of the data. Yet there may still be statistically
significant departures in terms of magnetic Prandtl number, but a greater range of $\Pm$ must be probed in global-scale
dynamo simulations to assess this, especially given the sparsity of the data at high Rossby number.  Nevertheless, none
of the fits are consistent with the energy-based scaling arguments, with the exception of a subpopulation in Figure
\ref{fig:scaling} and Figure \ref{fig:fit2}.  So, when considering the available simulation data, the buoyancy work
Rossby number scaling of $\mathrm{ME/KE} \propto \Ro^{-1/2}$ derived in \S\ref{sec:energybalance} and shown in Equation
(\ref{eqn:buoyancyscaling}) may be safely ruled out as a description of the magnetic energy scaling.

Within the context of the simulations exhibited in the full data set or Figure \ref{fig:scalingfull}, the rotation rates
employed lead to nearly three decades of coverage in Rossby number. In that figure, the force-based scaling derived in
\S\ref{sec:forcescaling} and given by Equation \ref{eqn:forcescaling} is depicted by the blue dashed curve, which does a
reasonable job of describing the nature of the superequipartition state for a given Rossby number. In particular, the
force balance studied here predicts a high Rossby number asymptotic regime.  This is captured in the data of
\citet{augustson16} in which there is a plateau of nearly equipartition states at high Rossby number, which points to a
balance between inertial and Lorentz forces in those simulations. This behavior is of note and should provide a
motivation for further investigating higher Rossby number dynamos. Surprisingly, the scaling law derived in
\S\ref{sec:energybalance} does not capture the behavior of this set of dynamos very well, in contrast to the many dynamo
simulations and data shown in \citet{christensen09} and \citet{christensen10} for which it performs well.  This could be
related to the fact that for large $\Pm$ the buoyancy work potentially has an additional $\mathrm{Ro}$ dependence
\citep{davidson13}. In particular, following \citet{brandenburg14}, the ratio of the dissipation rates can be described
as

\vspace{-0.25truein}
\begin{center}
  \begin{align}
    \frac{H_\nu}{H_\eta} = k \Pm^n,
  \end{align}
\end{center}

\noindent where $k$ and $n$ could be determined from a suite of direct numerical simulations. In particular, when
kinetic helicity is injected at the driving scale, \citet{brandenburg14} found that $k=7/10$ and $n=2/3$. Note that
their results have also considered rotating driven turbulence and found that this scaling is effectively independent of
the rotation rate, and thus the Rossby number.  However, other studies indicate that there may be a stronger rotational
influence \citep{plunian10}.  Since there is ambiguity in that scaling, let $k(\Ro)$ be an unknown function of the
Rossby number. Subsequently, assuming that $L_* \approx L_r$ in Equation \ref{eqn:fullnrg}, the buoyancy work $w_B$ per
unit mass can be described as

\vspace{-0.25truein}
\begin{center}
  \begin{align}
    w_B &= -\frac{\int \frac{dt}{\tau} dV \rho\vv\cnabla\Phi_{\mathrm{eff}}}{\int \frac{dt}{\tau}dV\rho} = \left(1+k(\mathrm{Ro}) \Pm^{n}\right)\frac{H_\eta}{M_{\mathrm{CZ}}},
  \end{align}
\end{center}

\noindent Such a scaling may provide a link between the moderate to high magnetic Prandtl number regime examined here to
the low magnetic Prandtl number regime assessed in \citet{davidson13}, which will be the focus of later work. It may
also possibly extend the analysis of \citet{davidson13} to baroclinic flows often found in stars rather than purely
barotropic flows that are more applicable to the geodynamo.

\section{Conclusions}

The characteristics of two descriptions of stellar and planetary dynamos have been discussed, where those dynamos are
distinguished by their magnetic Prandtl number and by the Rossby number of the convection that drives them. Scaling
relationships for the level of the partitioning of magnetic energy and kinetic energy have been derived from two
different principles. Particularly, there may be a shift in the kind of dynamo action taking place within stars that
possess a convective core and those that possess an exterior convective envelope if the atomic values of the
diffusivities provide the basis of the effective diffusivities, as demonstrated in Figure \ref{fig:prandtl}. The scaling
relationship between the magnetic and kinetic energies of such convective dynamos in turn provide an estimate of the rms
magnetic field strength in terms of the local rms velocity and density at a particular depth in a convective zone, which
will be of particular use to the stellar evolution community dealing with magnetic related instabilities and induced
diffusivities.

Given the atomic values of the diffusivities relevant to stellar interiors shown in Figure \ref{fig:prandtl}, the atomic
value of the magnetic Prandtl number shifts from a very low regime in low mass stars to a moderate $\Pm >1$ regime in
higher mass stars.  Thus, two scaling relationships for the ratio of the magnetic to kinetic energy appear to be the
most applicable for stellar and planetary dynamos: one for the high or moderate magnetic Prandtl number regime and
another for the low magnetic Prandtl number regime.  Within the context of the moderate to large magnetic Prandtl number
simulations in the current data set, the magnetic energy of the system scales as the kinetic energy multiplied by an
expression that depends on the inverse Rossby number plus an offset, which in turn depends upon the details of the
non-rotating system (e.g., on Reynolds, Rayleigh, and Prandtl numbers). On the other hand, for low magnetic Prandtl
number and fairly rapidly rotating systems, such as the geodynamo and rapidly rotating stars, another scaling
relationship that relies upon an energetic balance of buoyancy work and magnetic dissipation (as well as a force balance
between the buoyancy, Coriolis, and Lorentz forces) may be more applicable. When focused on in detail, this yields a
magnetic energy that scales as the kinetic energy multiplied by the inverse square root of the convective Rossby number.
Such a scaling relationship has been shown to be fairly robust \citep{davidson13} for some dynamo simulations.  With
currently available simulation data, there is no clear way to assess if there is a shift between these dynamo scaling
regimes.

However, there are some hints from dynamo simulations that there may be an influence of the magnetic Prandtl
number. Specifically, \citet{brun05,featherstone09}, and \citet{augustson16} find that the dynamos of intermediate and
high mass stars tend to have weak mean-field components and comparatively large non-axisymmetric magnetic energy,
whereas the bulk of lower mass star simulations appear to have stronger mean fields
\citep[e.g.,][]{augustson13,augustson15,brun17} as well as many other simulations such as those of
\citet{mabuchi15,guerrero16,varela16,strugarek18,viviani18}, and \citet{warnecke18}. Such behaviors comport with modern
dynamo theory \citep[e.g.,][]{brandenburg05}, where if the atomic values of the diffusivities as shown in Figure
\ref{fig:prandtl} influence the effective Pm, then it would be expected that low-mass stars should exhibit stronger mean
magnetic fields, which appear to be characteristic of low-Pm dynamos, and core convective dynamos of massive stars that
exhibit small-scale, high-Pm dynamo characteristics.

Indeed, additional work is needed to establish more robust scaling relationships that cover a greater range in both
magnetic Prandtl number and Rossby number, among other parameters. The current suite of simulations do not exhibit a
statistically significant Pm dependence in the data, which is likely due to the spread in the supercriticality of these
simulations and the restricted range of magnetic Prandtl numbers considered. Thus, numerical experiments should also
explore a larger range of Reynolds number and levels of supercriticality. Indeed, as in \citet{yadav16}, some authors
have already attempted to examine such an increased range of parameters for the geodynamo. Nevertheless, to be more
broadly applicable in stellar physics, there is a need to find scaling relationships that can bridge both the low and
high magnetic Prandtl number regimes that are shown to exist within main-sequence stars.

\section*{Acknowledgments}

The authors thank the anonymous referee for their helpful comments that improved the data analysis and strengthened the
conclusions. The authors further thank A. Strugarek and J. Varela for the use of their data, as well as S. Mathis for
extensive discussions and helpful advice.  K.~C. Augustson acknowledges support from the ERC SPIRE 647383
grant. A.~S. Brun acknowledges funding by ERC WHOLESUN 810218 grant, INSU/PNST, and CNES Solar Orbiter. J. Toomre was
partly supported by NASA through grants NNX14AB56G and NNX17AG22G.

\appendix

\section{Fitting Coefficients}\label{app1}

The mean value and standard deviation of the fitting coefficients, which are determined through the cross-validation
analysis, are given below in Table \ref{tab1} for the fits to the data shown in Figures
\ref{fig:scaling} and \ref{fig:scalingfull}. Figure \ref{fig:fit2} shows the histograms of the fitting coefficients for
the functions $f_0$ and $f_2$ in Figure \ref{fig:scaling}, which are the only functions of statistical relevance studied
here. Likewise, Figure \ref{fig:fit4} shows the same for Figure \ref{fig:scalingfull}.

  \begin{center}
    \begin{tabular}{cccccccc}
      \hline
      & & Figure 3  & & & & Figure 4 &\\
      \hline
      $f_i$ & Coefficient & Mean & Std. Dev. & $f_i$ & Coefficient & Mean & Std. Dev. \\
      \hline
      $f_0$ & & & & $f_0$ & & & \\
            & $a_0$ & $0.053$ & $0.004$ & & $a_0$ & $0.051$ & $0.007$ \\
            & $b_0$ & $0.062$ & $0.009$ & & $b_0$ & $0.06$ & $0.01$ \\
      $f_1$ & & & & $f_1$ & & & \\
            & $a_1$ & $0.16$ & $0.08$ & & $a_1$ & $0.2$ & $0.2$  \\
            & $b_1$ & $-0.1$ & $0.2$ & & $b_1$ & $-0.1$ & $0.2$ \\
            & $c_1$ & $-0.1$ & $0.1$ & & $c_1$ & $-0.2$ & $0.2$ \\
      $f_2$ & & & & $f_2$ & & & \\
            & $a_2$ & $0.04$ & $0.02$ & & $a_2$ & $0.046$ & $0.008$ \\
            & $b_2$ & $0.08$ & $0.07$ & & $b_2$ & $0.09$ & $0.03$ \\
            & $c_2$ & $-1.0$ & $0.4$ & & $c_2$ & $-0.9$ & $0.2$ \\
      $f_3$ & & & & $f_3$ & & & \\
            & $a_3$ & $0.03$ & $0.04$ & & $a_3$ & $0.06$ & $0.08$  \\
            & $b_3$ & $0.02$ & $0.05$ & & $b_3$ & $0.01$ & $0.03$ \\
            & $c_3$ & $0.1$ & $0.3$ & & $c_3$ & $0.1$ & $0.3$ \\
            & $d_3$ & $0$ & $3$ & & $d_3$ & $0.2$ & $0.3$ \\
            & $e_3$ & $-1$ & $1$ & & $e_3$ & $-1$ & $1$ \\
            & $f_3$ & $0.01$ & $0.04$ & & $f_3$ & $0.1$ & $0.2$ \\
            & $g_3$ & $-2$ & $3$ & & $g_3$ & $-1$ & $2$ \\
            & $h_3$ & $1$ & $1$ & & $h_3$ & $0.0$ & $0.1$ \\
     \hline
    \end{tabular}
    \captionof{table}{Fit parameters for the data of Figures \ref{fig:scaling} (left column) and \ref{fig:scalingfull} (right
      column), where only significant digits are shown.}\label{tab1}
  \end{center}

The goodness of fit results for the various fits and measures as discussed above are provided in Table \ref{tab2}.

  \begin{center}
    \begin{tabular}{ccccc}
      \hline
      \hline
      Figure and $f_j$ & $\chi_{\mathrm{CV},j}^2$ & $\chi_{\mathrm{rel, CV},j}$ & $\chi_{j}^2$ & $\chi_{\mathrm{rel},j}$\\
      \hline
      3 & & & &\\
      $f_0$ & $1.06$ & $9.03$ & $1.13$ & $9.20$ \\
      $f_1$ & $46.9$ & $11.5$ & $2.07$ & $13.8$ \\
      $f_2$ & $14.5$ & $6.48$ & $1.28$ & $9.67$ \\
      $f_3$ & $157$ & $12000$ & $53.7$ & $269000$ \\
      \hline
      4 & & & &\\
      $f_0$ & $1.06$ & $5.99$ & $1.07$ & $5.72$ \\
      $f_1$ & $6.81$ & $23.23$ & $4.38$ & $12.1$ \\
      $f_2$ & $1.04$ & $6.13$ & $1.08$ & $6.57$ \\
      $f_3$ & $4.57$ & $60.1$ & $5.40$ & $33.5$ \\ 
      \hline
    \end{tabular}
    \captionof{table}{Goodness of fit parameters for the data of Figures \ref{fig:scaling} (upper row) and \ref{fig:scalingfull} (lower
      row), indicating that only the model functions $f_0$ and $f_2$ are statistically significant fits to the data.}\label{tab2}
  \end{center}
  
\section{Atomically-Valued Diffusivities and Fluid Numbers}\label{app2}

It is useful to quantify the specific regimes in which different models of convective dynamos are most applicable.  To
do so, consider a fully-resolved convective dynamo and its associated dynamics, wherein the dissipation of the energy
injected into the system is governed by the atomic values of the diffusivities.  Using Braginskii plasma
diffusivities \citep{braginskii65} and a set of MESA stellar models with a solar-like metallicity to obtain the density
and temperature profiles \citep{paxton11}, one can define the average expected atomic magnetic Prandtl number in
either the convective core of a massive star or the convective envelope of a lower mass star.  In what follows, $\nu$ is
the kinematic viscosity, $\eta$ is the magnetic diffusivity, $\kappa_{\mathrm{cond}}$ is the electron thermal
diffusivity, $\kappa_{\mathrm{rad}}$ is the radiative thermal diffusivity. These quantities are computed under the
assumptions that charge neutrality holds and that the temperatures are not excessively high, so that the Coulomb
logarithm is well-defined. In the charge-neutral regime, the magnetic diffusivity happens to be density independent
because the electron collision time scales as the inverse power of ion density and the conductivity is proportional to
the electron density times the electron collision time.

\vspace{-0.25truein}
\begin{center}
  \begin{align}
    \nu = \left[2.21 \times 10^{-15} Z^{-3} \mu^{1/2} + 2.77 \times 10^{-17}\right] \lambda^{-1} \rho^{-1} T^{5/2},
  \end{align}
\end{center}

\noindent which has units of $\mathrm{cm^2\, s^{-1}}$, and where $\lambda$ is the Coulomb logarithm (defined below),
$\mu$ is the mean atomic mass (and it is unitless since $\mu$ is the ratio of the ion mass to the proton mass), $\rho$
is the density in $\mathrm{g\, cm^{-3}}$, $T$ is the temperature in $K$, and $Z$ is the mean atomic charge (unitless),
which is defined as $Z=\sum_{s=1}^{N_{S}} Z_s(T,P) n_s/n_i$ and where $n_i = \sum_{s=1}^{N_{S}} n_s$.

\vspace{-0.25truein}
\begin{center}
  \begin{align}
    \eta = 1.02 \times 10^{12} (2.86 Z - 0.90) \lambda T^{-3/2},
  \end{align}
\end{center}

\noindent which also has units of $\mathrm{cm^2\, s^{-1}}$. The thermal diffusivity arising from thermal conductivity is

\vspace{-0.25truein}
\begin{center}
  \begin{align}
    \kappa_{\mathrm{cond}} &= \Big\{7.41 \times 10^{-7} Z^{-3} \mu^{-1/2} 
                             + 7.20 \times 10^{-6} \tanh\left[0.13 \left(1+Z\right)\right]\Big\}\lambda^{-1} T^{5/2}.
  \end{align}
\end{center}

\noindent However, it is necessary to include the radiative contribution to the thermal conductivity:

\vspace{-0.25truein}
\begin{center}
  \begin{align}
    \kappa = \kappa_{\mathrm{cond}} + \kappa_{\mathrm{rad}},
  \end{align}
\end{center}

\noindent with

\vspace{-0.25truein}
\begin{center}
  \begin{align}
    \kappa_{\mathrm{rad}} = \frac{4 a c T^{3}}{3 c_P \rho^{2} \sigma},
 \end{align}
\end{center}
    
\noindent where $a$ is the radiation constant, $c$ the speed of light, $c_P$ is the pressure specific heat, and $\sigma$
is the opacity.  With those in hand, one can then compute the thermal and magnetic Prandtl numbers as well as the
magnetic Reynolds number as necessary. The Coulomb logarithm is quite complex as it depends on the number of included
species and on the details of their Landau scattering integrals.  However, it only varies between roughly 1 and 30
within the context of stars.  So, for now, set $\lambda = 15$ (which is an average value for densities between
$10^{-10}$ and $10^{10}$ $\mathrm{g cm^{-3}}$, and temperatures between $10^3$ and $10^{10}$ $\mathrm{K}$). Therefore,
the thermal Prandtl number $\Prt$ and magnetic Prandtl number $\Pm$ scale as

\vspace{-0.25truein}
\begin{center}
  \begin{align}
    \Prt &= \frac{\nu}{(\kappa_{\mathrm{cond}}+\kappa_{\mathrm{rad}})}, \qquad \Pm = \frac{\nu}{\eta}.
  \end{align}
\end{center}

\bibliography{scaling}

\end{document}